\documentclass[aps,twocolumn,groupedaddress,prb,floatfix,showpacs]{revtex4}
\usepackage{graphicx}
\usepackage{dcolumn}
\usepackage{bm}
\usepackage{amsmath}
\usepackage{verbatim}

\begin{document}

\title{Graphene on Insulating Oxide Substrates:  Role of Surface Dangling States}

\author{Priyamvada Jadaun}
\author{Sanjay K. Banerjee}
\author{Bhagawan Sahu}
\email{brsahu@physics.utexas.edu}
\affiliation{
Microelectronics Research Center, The University of Texas at Austin, Austin Texas 78758}
\date{\today}

\begin{abstract}
We study the effect of insulating oxide substrates on the energy band structure of monolayer and bilayer graphene using a \textit{first principles} density functional based electronic structure method and a local exchange correlation approximation. We consider two crystalline substrates, SiO$_{2}$ (or $\alpha$-quartz) and Al$_{2}$O$_{3}$ ($\alpha$-alumina or sapphire), each with two surface terminations. We focus on the role of substrate surface dangling states and their passivation in perturbing the linear energy spectrum of graphene. On non-passivated surface terminations, with the relaxation of top surface layers, only Si-terminated quartz retains the linear band structure of graphene due to relatively large equilibrium separation from the graphene layer whereas the other three surface terminations considerably distort it. However, without relaxations of the top surface layer atoms, linear bands appear in the electronic spectrum but with the Dirac point shifted away from the Fermi level.  Interestingly, with a second carbon layer on non-passivated oxygen terminated Quartz, with top surface layers relaxation, graphene features appear in the spectrum but sapphire with both surface terminations shows perturbed features even with two carbon layers. By passivating the surface dangling states with hydrogen atoms and without top layer atomic relaxations, the electron-hole symmetry occurs exactly at the Fermi level. This suggests that surface dangling states play a less important role than the atomic relaxations of the top surface layers in distorting the linear spectrum. In all cases we find that the first layer of graphene forms ripples, much like in suspended graphene, but the strength of rippling is found to be weaker probably due to the presence of the substrate. We discuss the energetics of both passivated and non-passivated surface terminations with a top graphene layer.

\end{abstract}

\maketitle

\section{Introduction}

The excitement in graphene research, due to the realization of {\it table-top} high-energy physics experiments and the promise to replace silicon in future semiconductor chips, is certainly overwhelming \cite{geim}. Advancement in understanding the fundamental physics of electron and hole transport in it\cite{neto} and using graphene for spintronics \cite{vanwees} is critical to the realization of carbon-based electronics. However, to make a transition from graphene science to engineering and finally to a viable technology, it is crucial to understand the interaction of graphene with external parameters such as substrates\cite{rod}, contacts for measurements\cite{goldhaber}, and the role of high dielectric constant oxides\cite{tutuc} in addition to studying the effects of intermediate species which are present in a particular process flow of graphene device fabrication. These parameters can pose a fundamental challenge to the ultimate realization of carbon electronics. In this article, we report the role of one of these external parameters, namely the substrates, in changing the electronic structure of graphene using a density functional based electronic structure method\cite{kresse, kresse1} and a local approximation to the exchange-correlation potential (LDA)\cite{ca}. We use two insulating crystalline substrates, SiO$_{2}$ (or quartz) and Al$_{2}$O$_{3}$ (or sapphire), each with two surface terminations in our study. These are widely used substrates in experiments that use exfoliated graphene. We consider surface dangling state passivation with hydrogen atoms and atomic relaxations and focus on their role in perturbing the graphene electronic spectrum. We discuss the energetics and the energy band structures that result from the interaction of graphene with these surface terminations. 

Experimentally, there are studies which attempt to understand the atomic structure of graphene on insulating substrates\cite{fuhrer}, using nanometer scale microscopic techniques such as scanning tunneling and atomic force microscopies. In addition, using Raman spectroscopy, the role of substrates on phonon dispersions is also reported which indirectly hints at the change in the electronic spectrum of graphene due to the presence of substrates\cite{balandin}. Moroever, we are aware of two recent density functional studies of monolayer graphene on a crystalline SiO$_{2}$ substrate\cite{theory, nayak}. Due to different and insufficient structural details in those reports, we could not compare our results with the conclusions reached in References 12 and 13.

\begin{figure}[ht]
\includegraphics[width=1\linewidth]{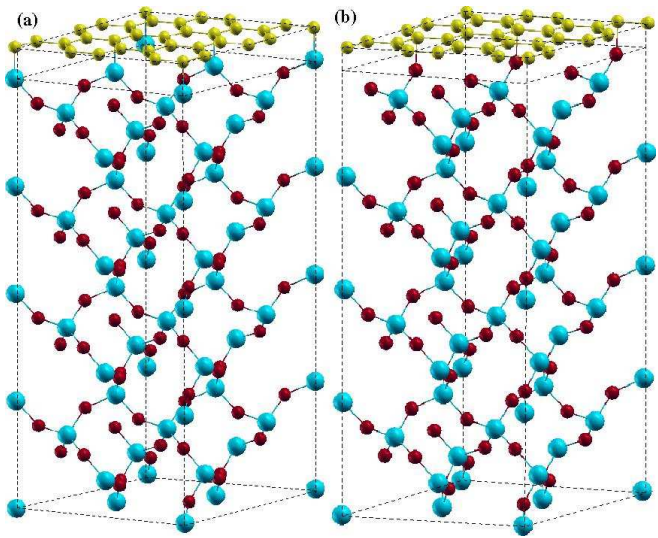}
\caption{ (Color online) Schematic illustration of the supercell structure of monolayer graphene on the top of (a) Si and (b) Oxygen-terminated quartz. Four unit cells of quartz are shown with Si atoms in blue, oxygen atoms in red and carbon atoms in yellow. Each Si atom is surrounded by four oxygen atoms.The supercell structure of graphene on the top of both Al and oxygen terminated sapphire is similar except that the successive Al and oxygen layers are stacked vertically at an angle. Hence we do not show them here.}
\label{fig:fig1}
\end{figure}         

Our paper is organized as follows. In section II, we first discuss the computational method and the convergence parameters used for this study  followed by the motivation and the procedure for building the surface models from their corresponding bulk counterparts. Our results of the effects of dangling state and its passivation on the linear spectrum of graphene and the role of atomic relaxation is discussed in section III. We describe the band structure with the help of orbital and atom projected densities of states. We then present the energetics of graphene on both non-passivated and passivated surface terminations. Finally, we summarize our results and present our conclusions.

\section{Computational Method and Surface Models} 
This section addresses the details of the computational method we used followed by the procedure we adopted to obtain the surface models of quartz and sapphire from their bulk counterparts and the convergence parameters used for this study. We used a plane-wave based electronic structure method with local density approximation (LDA)\cite{kresse} for exchange and correlation and the projector augmented plane-wave potential for electron-ion interaction\cite{kresse1}. The bulk structures of both the quartz and sapphire are consistent with those available in literature \cite{poindexter} and the surface models built from the bulk structures conform to the widely accepted $\alpha$-quartz and $\alpha$-alumina structures \cite{somorjai}. We first optimized the lattice parameters of bulk crystalline quartz and sapphire. For both the bulk substrates, we generated hexagonal unit cells from the original rhombohedral cells and these bulk phases form layered structures (alternating cation and anion layers). The unit cell of quartz contains 27 atoms with 3 Si planes and 6 oxygen planes, each plane containing 3 atoms, whereas the unit cell of sapphire contains 30 atoms with 4 aluminum (Al) planes and 6 oxygen planes again each plane containing 3 atoms. We used a 7 $\times$ 7 $\times$ 5 {\bf k}-point mesh in the hexagonal Brillouin zone (BZ) and a 612 eV kinetic energy cut-off. The results were carefully checked with respect to a larger {\bf k}-point set and higher energy cut-offs. 

\begin{figure}[ht]
\scalebox{0.4}{\includegraphics{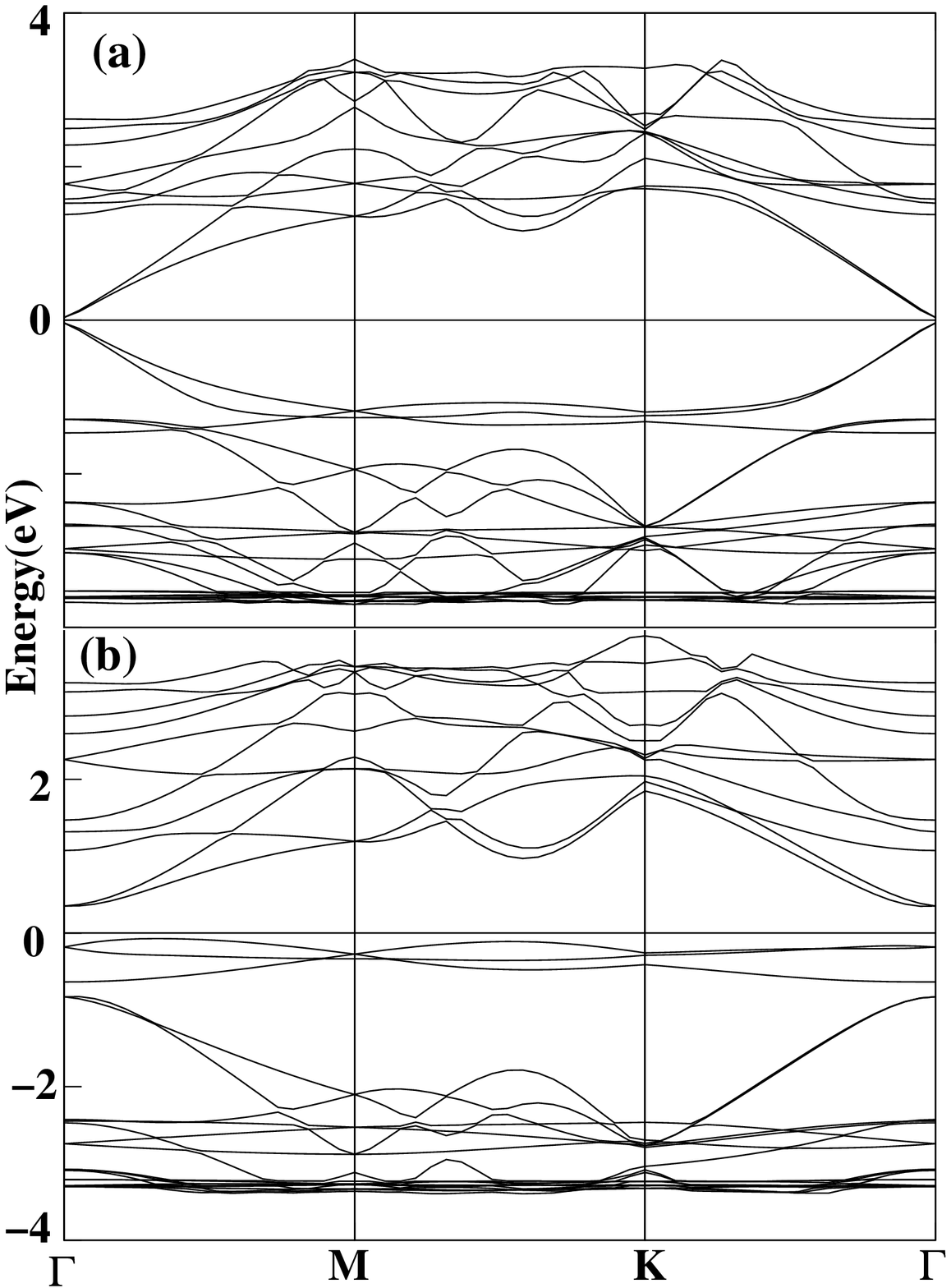}}
\caption{Energy band structures of monolayer graphene on Si-terminated quartz, at high-symmetry points in the supercell Brillouin zone, at the interlayer distances of (a) 3 {\rm \AA} and (b) 2.5 {\rm \AA}. The atoms in the supercell were not relaxed and the top surface was not passivated. The interlayer distance correponds to the location of graphene plane from the topmost Si-plane in the Si-terminated quartz substrate. The Fermi energy is set at zero. The origin of occurance of linear bands at the $\bf \Gamma$-point instead of {\bf K}-point of the supercell is explained in the text.}
\label{fig:fig2}
\end{figure}

We find that the optimized lattice constants of quartz and sapphire are close to the experimental values (Table I). The in-plane and out-of-plane lattice constants for quartz and sapphire differ, from the experimental values, by less than 1 $\%$. Using these optimized lattice parameters, we constructed surface models of both quartz and sapphire as follows. Four bulk unit cells were stacked along the {\it c}-direction (which corresponds to a thin-film thickness of about 22 {\rm \AA}) and we find that 6 $\times$ d$_{C-C}$ graphene, containing a total of 24 carbon (C) atoms (where d$_{C-C}$ =1.42 {\rm \AA}) is nearly commensurate with the hexagonal surface of the substrates (Fig. \ref{fig:fig1}). The lattice mismatch of the quartz and sapphire terminations with graphene is calculated to be 0.19 $\%$ and 0.42 $\%$ respectively. We note that the commensurability of graphene with the underlying substrate is not necessary for our study since it is not focussed on epitaxial growth of graphene on SiC and other metal substrates. Therefore, the lattice mismatch values mentioned here serve only as an initial guideline in assessing the degree of distortion of strictly two-dimensional graphene in the presence of the substrates. In fact, we observe rippling of graphene layers on each of the surface terminations in the final relaxed structures. We note that the atomic-scale roughness simulated here by relaxing top few layers of substrates does not resemble the roughness present on amorphous SiO$_2$ (or {\it a}-SiO$_2$) and Al$_2$O$_3$ (or {\it a}-Al$_2$O$_3$) surfaces that are used in experiments involving graphene. Simulating roughness of amorphous surfaces, using DFT based electronic structure method, will increase the computational burden significantly due to the requirement of large supercell size which is necessary to achieve roughness at the macroscopic scale. However, in the surface models adopted here, in their final equilibrium configurations, some roughness is present at the atomic scale. Periodic boundary conditions were enforced along the surface directions whereas a vacuum size of 10 {\rm \AA} was used along the {\it c} direction to enable periodic slab calculations. The silicon (Si) dangling states at the bottom of the supercell were saturated with hydrogen atoms. We fixed the supercell lattice parameters in all cases and only the atoms in the planes of the top two unit-cells of the substrate and the atoms in the graphene planes were allowed to relax. We used the same energy cut-off as in the bulk calculations but the {\bf k}-point mesh in the BZ was chosen to be 7 $\times$ 7 $\times$ 1. For atomic relaxations, the total energy was assumed to have converged when all the components of the Hellman-Feynman forces were smaller than 0.01 eV/{\rm \AA}.

\begin{figure}[ht]
\scalebox{0.4}{\includegraphics{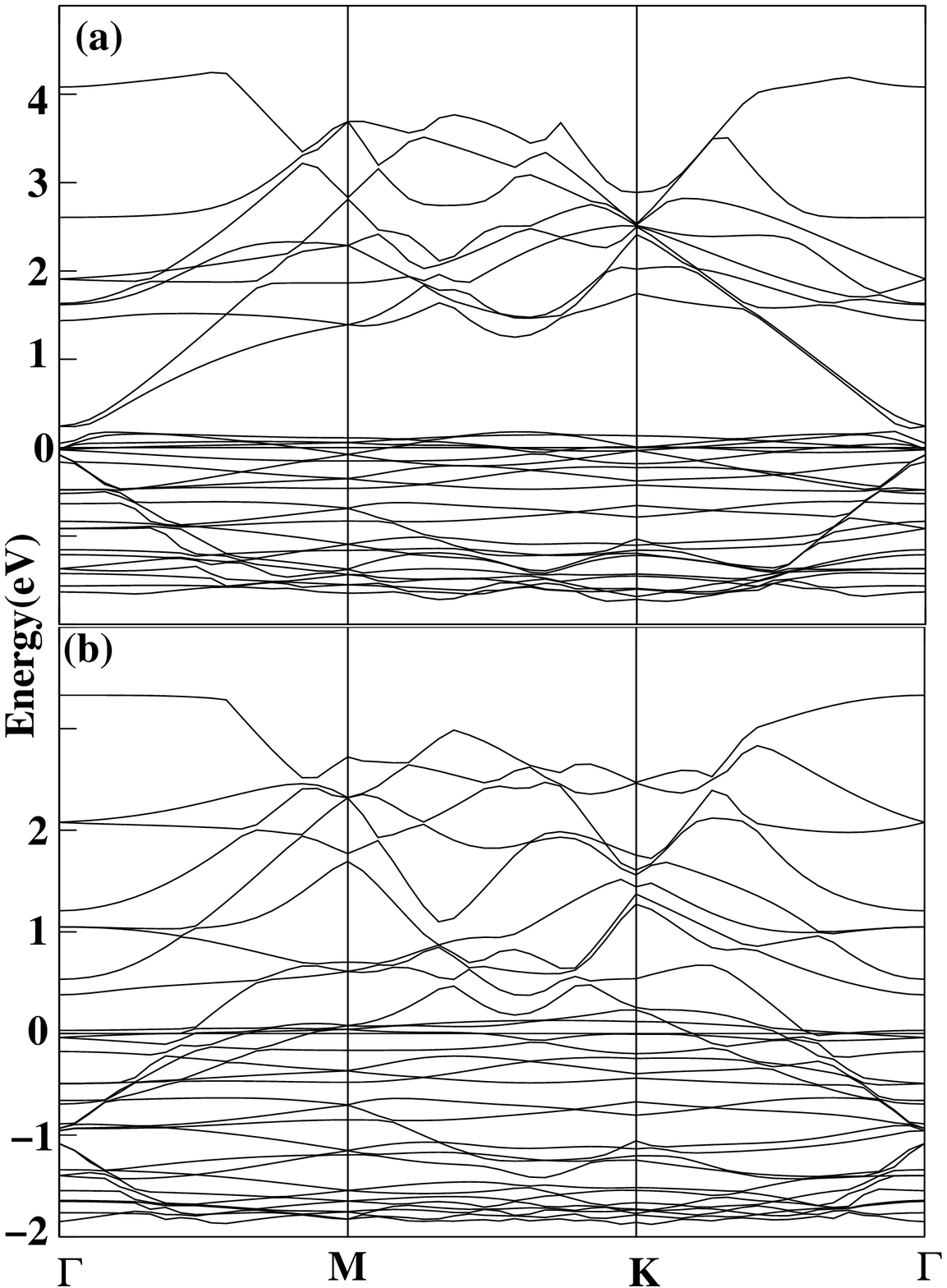}}
\caption{Energy band structures of monolayer graphene on Al-terminated sapphire, along high-symmetry points in supercell Brillouin zone, at interlayer distances of (a) 4.0 {\rm \AA} and (b) 2.7 {\rm \AA}. The atoms in the supercell were not relaxed and the top surface was not passivated. The interlayer distance correponds to the location of the graphene plane from the topmost Al-plane in Al-terminated sapphire substrate. The Fermi energy is set at zero.}
\label{fig:fig3}
\end{figure}

\section{Energy Band dispersions, Densities of states and Energetics}

In this section, we discuss the role of dangling states, their passivation and the role of atomic relaxations on the electronic band structure of graphene on the surface terminations we considered here. We note that both the surface terminations of quartz possess large number of dangling states whereas in sapphire, by construction of the surface models from bulk, the number of nearest neighbors of both Al and oxygen is found to be optimal. Therefore, for passivation studies, only quartz surfaces were considered. We passivated the dangling states with hydrogen atoms. There are four different possibilities that emerge from considering whether surface dangling states are passivated or not and whether or not atomic relaxations are performed. 

We first discuss Si-terminated quartz. We did not passivate the dangling states and did not relax the atomic positions. To get a rough estimate of the interlayer separation at which graphene electronic structure is minimally perturbed, we placed the graphene layer at various distances from the top substrate layer. The choice of these distances and distances chosen for other surface terminations, are arbitrary. We chose bilayer graphene interlayer distance (which is 3.34 {\rm \AA}) as a reference. At a distance of 3 {\rm \AA} and above, from the top Si surface, graphene retains the linear spectrum but at a smaller distance (of 2.5 {\rm \AA}), the linear bands are perturbed (Fig. \ref{fig:fig2}(a)and (b)). The occurance of the linear spectrum at the $\bf \Gamma$-point of the supercell BZ instead of the {\bf K}-point is due to crystal symmetry of the supercell and its relation to the monolayer graphene lattice symmetry. The origin of such shifting of the location of the electron-hole symmetry from the {\bf K}-point to the $\bf \Gamma$-point is seen in calculations involving sub-monolayer alkali metal adsorption on graphene surfaces\cite{farjam}. We discuss the origin of this shift in the Appendix section. On relaxing the atomic positions of the top few layers of the quartz substrate and graphene, we get an equlibrium distance of 3 {\rm \AA} between the top substrate layer and graphene due to which the perturbation to electronic spectrum is minimal, as in the non-relaxed case (Figure not shown). 

We now discuss Al-terminated sapphire and its interaction with graphene. For the non-passivated surface without atomic relaxations, the dispersion curve shows the dangling states at the Fermi level are resonance with the carbon $p_z$ orbital at interlayer distances as high as 4 {\rm \AA} and as low as 2.7 {\rm \AA} (Fig. \ref{fig:fig3}(a) and (b)). However, with atomic relaxations, we get an equlibrium separation of 2.7 {\rm \AA} and a similar energy dispersion curves (Figures now shown).   

\begin{figure}[ht]
\scalebox{0.4}{\includegraphics{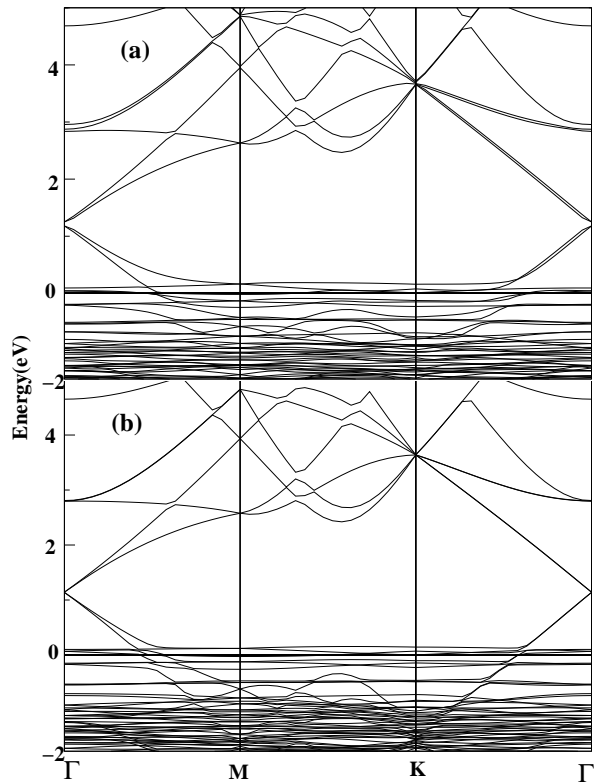}}
\caption{Energy band structures of monolayer graphene on oxygen-terminated quartz, at a interlayer distance of (a) 1.76 {\rm \AA} and (b) 2.5 {\rm \AA}.  The atoms in the supercell were not relaxed and the top surface was not passivated. The interlayer distance correponds to the location of the graphene plane from the topmost oxygen plane in oxygen-terminated quartz substrate. The Fermi energy is set at zero.}
\label{fig:fig4}
\end{figure}

\begin{figure}
\scalebox{0.4}{\includegraphics{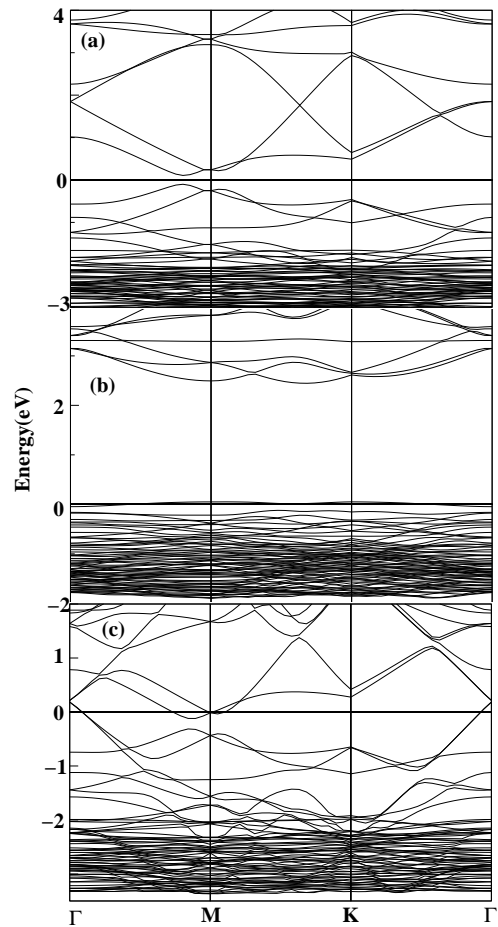}}
\caption{Energy band structures of monolayer graphene on (a) oxygen-terminated quartz, (b) oxygen-terminated sapphire, along high-symmetry points in supercell Brillouin zone, computed at the respective equilibrium separations shown in Table I and (c) two carbon layers on the top of the oxygen-terminated quartz seems to restore the linear behavior at the Fermi level. The atoms in the supercell were relaxed but the top surface was not passivated. The Fermi energy is set at zero.}
\label{fig:fig5}
\end{figure}

The situation is similar in both oxygen-terminated quartz and sapphire. On non-passivated oxygen-terminated quartz without atomic relaxations, at distances of 1.76 {\rm \AA} and 2.5 {\rm \AA}, graphene features appear in the band structure (Fig. \ref{fig:fig4}(a) and (b)) but the electron-hole symmetry is located above the Fermi level but an energy gap opens up for the case of 2.5 {\rm \AA}. Opening of an energy gap may be due to the breaking of sub-lattice symmetry in graphene as observed in epitaxial growth of graphene on SiC substrate\cite{lanzara} but is most likely due to approximations in DFT. On relaxing the atomic positions, oxygen-terminated surfaces in both quartz and sapphire, maintain an equilibrirm separation which is quite close to the graphene layer. As a result, we find a strong perturbation to the linear spectrum of graphene (Fig. \ref{fig:fig5}(a) and (b)).  

When we add a second carbon layer on oxygen-terminated quartz and relax the atomic positions, we recover linear spectrum of graphene (Fig. \ref{fig:fig5}(c)). This hints at the existence of a buffer layer in studies of graphene on oxygen-terminated quartz. However, our calculations suggest more than two carbon layers are needed to recover graphene linear energy spectrum in case of Al- and oxygen-terminated sapphire.              

These equlibrium distances, for each non-passivated surface terminations with atomic relaxations, are listed in Table I. It also lists interatomic distances between the Carbon and the corresponding atom of the surface terminations in the bulk or molecular phase. Since all the equilibrium distances are larger than that in the corresponding interatomic distances in the bulk or molecular phases, it suggests that graphene does not form a bulk- or molecular-like phase with the surface terminations used in this study. 

\begin{figure}
\scalebox{0.4}{\includegraphics{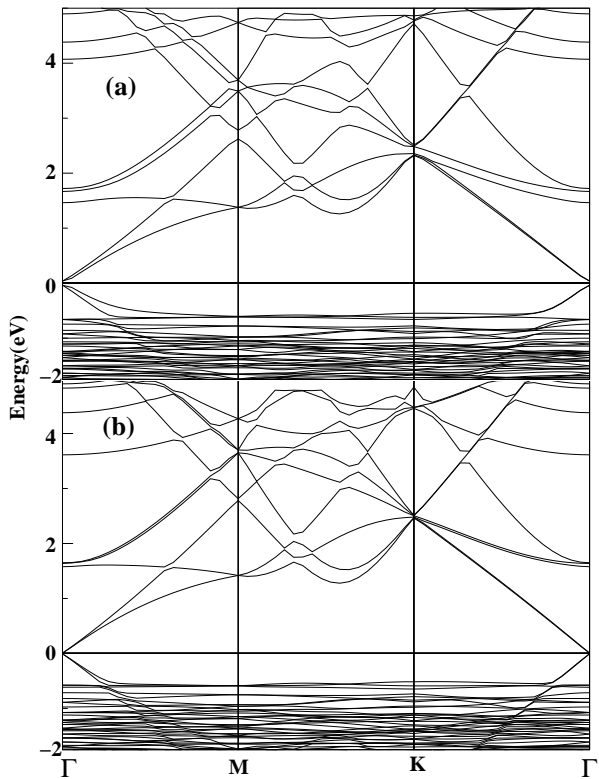}}
\caption{ (Color online) Same as Figure 4 but now for passivated surface of oxygen terminated quartz.}  
\label{fig:fig6}
\end{figure}

We now discuss the effect of passivation of dangling states on the band structure of graphene. Since graphene on non-passivated Si-terminated quartz already shows the a linear spectrum, we considered passivating only the dangling states of oxygen-terminated quartz. We did not relax the atomic positions. Linear bands seem to appear above the Fermi level for distances below 2 {\rm \AA} but right at the Fermi level in case of distances above 2 {\rm \AA} (Fig. \ref{fig:fig6}(a) and (b)). It is likely that relaxing the atomic positions may retain the graphene features but with a different equilibrium separation than the non-passivated surface. We did not consider this case explicitly in our calculations.

Our results suggest that presence or absence of dangling states is not as effective in distorting the linear band structure as the atomic relaxations. It is the equilibrium distance that dictates the survival of graphene features in the dispersion spectrum. However, a case in which relaxations of passivated substrate and graphene provide a favorable equilbrium distance for retention of the graphene features, is not ruled out.   

\begin{figure}[ht]
\scalebox{0.4}{\includegraphics{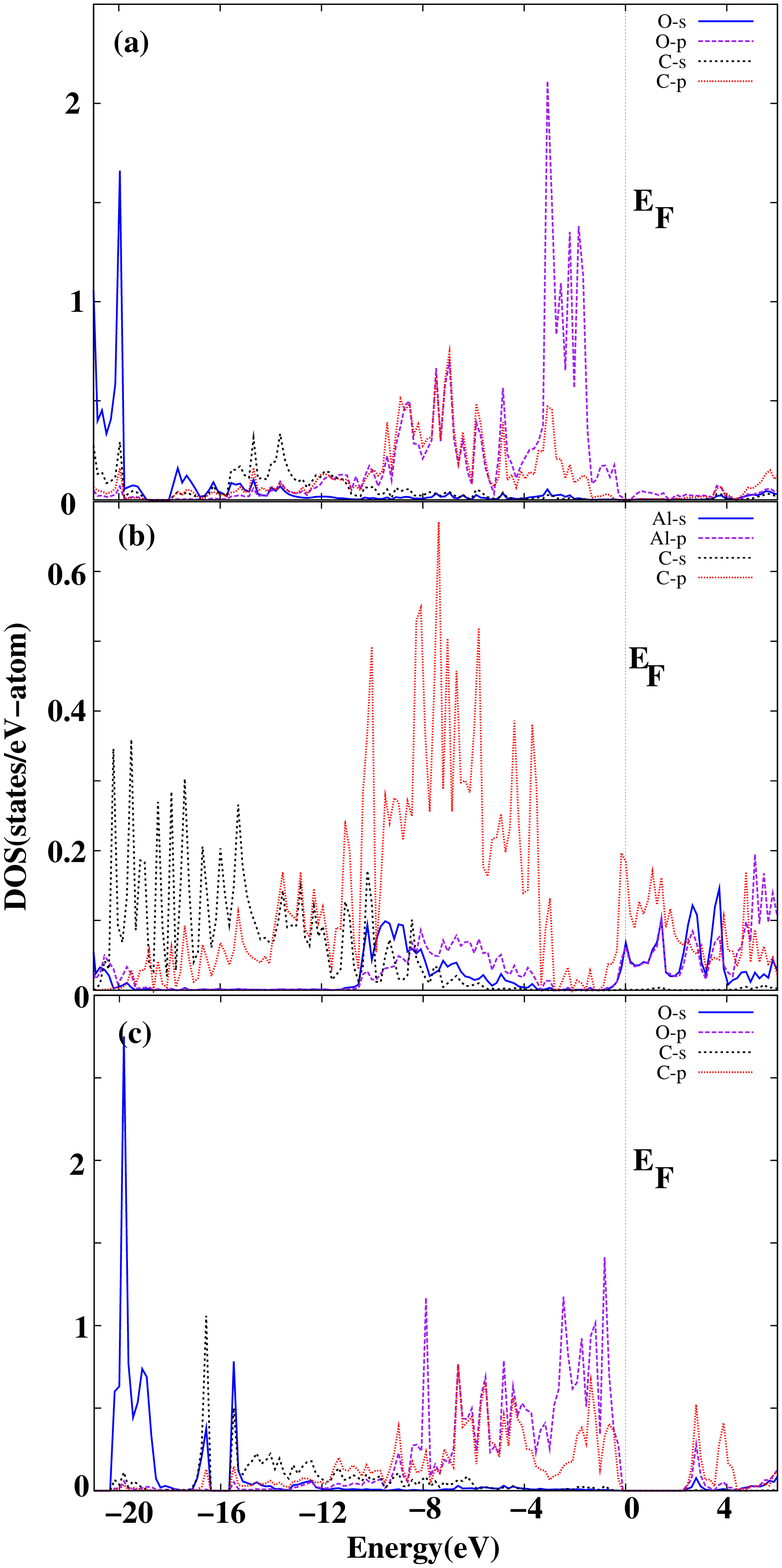}}
\caption{ (Color online) Atom and orbital projected density of states for a C atom in monolayer graphene and the atoms of the topmost planes in (a) oxygen-terminated quartz (b) Al-terminated sapphire  and (c) oxygen-terminated sapphire substrates computed at the equilibrium separations shown in Table I. The atoms in the supercell were relaxed but the top surface was not passivated. The Fermi energy is set at zero.}  
\label{fig:fig7}
\end{figure}

To understand the perturbations to the linear spectrum of graphene in case of non-passivated relaxed surface, we plot atom and orbital projected densities of states (or DOS) (Fig. \ref{fig:fig7}). The top panel (a) shows DOS for oxygen terminated quartz, projected on to {\it s}- and {\it p}-states of oxygen atom of the top surface layer and carbon. We see a strong hybridization between oxygen-{\it p} and C-{\it p} orbitals in the vicinity of the Fermi level. A similar conclusion was reached in a recent DFT-based calculation on monolayer graphene on oxygen-terminated quartz\cite{nayak}. Such resonance structures are also seen in the DOS of both Al- and oxygen-terminated sapphire (panels (b) and (c)). This explains why graphene $\pi$-orbitals, after atomic relaxations, cannot retain their linear dispersion in the presence of non-passivated surfaces. We also estimated average height fluctuations of two-dimensional graphene due to the presence of substrate and the atomic relaxations of the top surface layers. In all cases, we find that the deviations are $\sim$ 0.05 {\rm \AA} which is small compared to suspended graphene\cite{kats}.
  
Table I shows the binding energy values for graphene on the four surface terminations considered in this study. In the most relevant case i.e. no surface passivation but with atomic relaxations, these values hint at non-bonding nature of graphene to the underlying oxide substrates. The binding energy values are obtained by using the following definition.

\begin{equation}
 {\rm {\it E}^{bind} = {\it E}(supercell)- {\it E}(Gr) - {\it E}(substrate)}
\end{equation}

where {\it E}(supercell) denotes the total energy of the supercell containing the substrate and a graphene layer. {\it E}(Gr) and {\it E}(substrate) denote, respectively, the total energies of isolated graphene and isolated substrate in the same supercell set-up, with the same energy cut-off and {\bf k}-point mesh as that of the combined graphene and substrate calculations.

\begin{table}[ht]
\caption{
Lattice parameters (in {\rm \AA}) of bulk quartz and sapphire, average interplanar distances (in {\rm \AA}) of graphene from the four underlying surface terminations and their binding energies (in eV/atom), in case of non-passivated but relaxed surfaces. The numbers in parenthesis are the out-of-plane lattice constants of the bulk phases and those in square parenthesis are interatomic distances between Carbon and the corresponding atom of the surface terminations in the bulk or molecular phase. It should be noted that, for isolated substrate calculations, we passivated the top layer with hydrogen atoms to keep the supercell non-magnetic.  
}
\label{tab1}
\begin{ruledtabular}
\begin{tabular}{cccccccc}
                       &  Lattice Paramters & d(C-x)                & E$^{bind}$       \\ 
                       &                     & (x = Si,Al,O)         &                    \\ \hline
Bulk quartz            &                    &                       &                   \\
 This work             &  4.914 (5.408)     &                       &                   \\
 Expt.\footnote{Reference 14, 16}    &  4.913 (5.405)     &                   &                   \\
Si-terminated           &                    &    3.0 (1.89)\footnote{Reference 17}         & 10.061            \\
Oxygen-terminated           &                    &    1.76(1.3)\footnote{Reference 18}         & 0.581            \\
                       &                    &                       &                     \\ \hline
Bulk sapphire          &                    &                       &                     \\
This work              & 4.907 (4.908)      &                       &                     \\
Expt.\footnote{Reference 15, 19}     & 4.943 (4.907) &                       &                     \\
Al-terminated          &                    &   2.7 (1.89-2.19)\footnote{Reference 20}     &   1.017      \\
Oxygen-terminated          &                    &   2.15                &   0.689              \\
\end{tabular}
\end{ruledtabular}
\end{table}

However, we find that passivated but non-relaxed surfaces are energetically more favorable compared to both non-passivated non-relaxed as well as non-passivated relaxed surfaces. This suggests an existence of a metastable configuration of graphene in presence of the non-passivated surfaces (both with and without atomic relaxations). Of course, passivated and relaxed surfaces, which we did not explicitly calculate here,  can be lower in energy than any of these cases.

\section{Summary and Conclusion} 

In summary, we have studied, using a {\it first principles} DFT method, the effect of two crystalline insulating substrates, quartz and sapphire, on the electronic structure of monolayer graphene. We considered the effect of surface passivation and atomic relaxations on the linear spectrum of graphene. Dangling states seems to be ineffective in distorting the linear band structure of graphene except shifting the spectrum above Fermi level. It is the atomic relxations which dictate the equlibrium separation between graphene and the topmost surface layer and this distance decides the strength of perturbation on linear bands. Si-terminated $\alpha$-quartz retains the graphene band structure on non-passivated surface even with atomic relaxations. This leaves oxygen-terminated quartz surface where atomic relaxations play an important role. Both Al- and oxygen-terminated sapphire perturb the linear bands whether or not their dangling states are passivated and whether or not atomic relaxations are performed. Two carbon layers are necessary to recover the linear band structure of graphene on the top of oxygen-terminated quartz whereas our calculations hint at more than two carbon layers are needed in case of Al- and oxyen-terminated Sapphire. Energetically, graphene on non-passivating surface terminations is metastable. 

\acknowledgments

The authors acknowledge financial support from the DARPA-CERA and NRI-SWAN program. The authors acknowledge the allocation of computing time on NSF Teragrid machines {\it Ranger} (TG-DMR080016N) and {\it Lonestar} at Texas Advanced Computing Center.    

\appendix*
\section{}

Quartz and sapphire substrates and the graphene layer, all have a hexagonal unit cell. Let $\roarrow A_{1}$, $\roarrow A_{2}$ and $\roarrow A_{3}$ be the lattice vectors of the entire graphene and substrate system and $\roarrow B_{1}$, $\roarrow B_{2}$, and $\roarrow B_{3}$ be the correponding reciprocal lattice vectors. Similarily, let ($\roarrow a_{1}$, $\roarrow a_{2}$ , $\roarrow a_{3}$) and ($\roarrow b_{1}$, $\roarrow b_{2}$ and $\roarrow b_{3}$) be the triad of primitive and reciprocal lattice vectors of the graphene layer alone.

We note that the lattice structure of the graphene plus substrate system is a (2 $\sqrt(3)$ x 2 $\sqrt(3)$)R30$^o$ reconstruction of the lattice structure of the graphene layer alone.
Thus $A = 2 \sqrt(3) a$ where A is the lattice constant of the entire system and a is the lattice constant of the two-dimensional graphene layer. Since both unit cells are hexagonal this implies that the reciprocal lattice structure of the graphene layer on top is a ($2\sqrt(3)$ x $2\sqrt(3)$)R30$^o$ reconstruction of the reciprocal lattice structure of the graphene plus substrate system below. Taking $\roarrow a_{1}$ to be along $\hat x$,

\begin{equation}
 {\rm \roarrow a_{1} = {\it a} \hat x} 
\end{equation} 
\begin{equation}
 {\rm \roarrow a_{2} = {\it a} (\frac{1}{2} \hat x + \frac{\sqrt(3)}{2} \hat y)} 
\end{equation} 
\begin{equation}
 {\rm \roarrow A_{1} = 2 \sqrt(3) {\it a}(\frac{\sqrt(3)}{2} \hat x + \frac{1}{2} \hat y)} 
\end{equation} 
\begin{equation}
 {\rm \roarrow A_{2} = 2 \sqrt(3) {\it a} \hat y} 
\end{equation} 
\begin{equation}
 {\rm \roarrow A_{3} =  {\it c} \hat z} 
\end{equation}

\begin{figure}[ht]
\scalebox{0.4}{\includegraphics{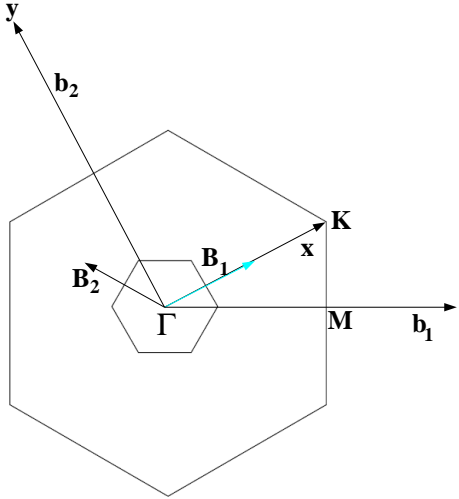}}
\caption{ (Color online) Schematics of supercell Brillouin zone (small) superposed on the graphene honeycomb Brillouin zone (large). The reciprocal lattice vectors are shown along with the overlap of $\bf \Gamma$ to {\bf K} vector of graphene with the {\bf B$_1$} vector of the supercell. The overlap is due to the symmetry reasons and is explained in the text.}   
\label{fig:fig8}
\end{figure}

The reciprocal lattice vectors are,
 
\begin{equation}
 {\rm \roarrow b_{1} = \frac{2 \pi}{\it a} (\hat x - \frac{1}{\sqrt(3)} \hat y)} 
\end{equation} 
\begin{equation}
 {\rm \roarrow b_{2} = \frac{2 \pi}{\it a} ( \frac{2}{\sqrt(3)} \hat y)} 
\end{equation} 
\begin{equation}
 {\rm \roarrow B_{1} = \frac{2 \pi}{ 2 \sqrt(3) \it a} ( \frac{2}{\sqrt(3)} \hat x)} 
\end{equation} 
\begin{equation}
 {\rm \roarrow B_{2} = \frac{2 \pi}{ 2 \sqrt(3) \it a} ( -\frac{1}{\sqrt(3)} \hat x + \hat y)}
\end{equation} 

Figure 8 shows the BZ set-up with these vectors using equations A.6 to A.9.

Now if we take the same $\Gamma$ point as the center of both the Brillouin zones and draw the reciprocal unit cell of both the structures we realize that the {\bf K } point of only the graphene layer lies on the reciprocal lattice vector $\roarrow B_{1}$ of the graphene plus substrate system (as shown in figure). This means that the {\bf K } point of graphene layer folds in by symmetry onto the $\Gamma$ point of the entire system.

\end{document}